\documentclass[10pt,journal,compsoc]{IEEEtran}
\ifCLASSOPTIONcompsoc
  \usepackage[nocompress]{cite}
\else
  \usepackage{cite}
\fi
\ifCLASSINFOpdf
\else
\fi
\usepackage{amsmath,amsfonts}
\usepackage{algorithmic}
\usepackage{textcomp}
\usepackage{multirow}
\usepackage{romannum}
\usepackage{xcolor}
\usepackage{tikz}
\usetikzlibrary{spy}
\usepackage{subcaption}
\usepackage{float}
\usetikzlibrary{shapes.geometric, arrows}
\usepackage{adjustbox}
\usepackage{tabularx}
\usepackage{makecell}
\usepackage{rotating}
\usepackage{adjustbox}
\usepackage{lscape}
\usepackage{cleveref}
\usepackage{array,collcell}

\begin{document}
%
\title{Transport Network, Graph, and Air Pollution}
%
%
%
%

\author{Nan~Xu
\IEEEcompsocitemizethanks{\IEEEcompsocthanksitem N. Xu is with the Department
of Infrastructure Engineering, University of Melbourne, VIC,
Australia, 3010.\protect\\
E-mail: naxu@student.unimelb.edu.au}
\thanks{Manuscript received February 17, 2025.}}

%
%

\markboth{Journal of \LaTeX\ Class Files,~Vol.~14, No.~8, August~2015}%
{Shell \MakeLowercase{\textit{et al.}}: Bare Demo of IEEEtran.cls for Computer Society Journals}
%



\IEEEtitleabstractindextext{%
\begin{abstract}
Air pollution can be studied in the urban structure regulated by transport networks. Transport networks can be studied as geometric and topological graph characteristics through designed models. Current studies do not offer a comprehensive view as limited models with insufficient features are examined. Our study finds geometric patterns of pollution-indicated transport networks through 0.3 million image interpretations of global cities. These are then described as part of 12 indices to investigate the network-pollution correlation. Strategies such as improved connectivity, more balanced road types and the avoidance of extreme clustering coefficient are identified as beneficial for alleviated pollution. As a graph-only study, it informs superior urban planning by separating the impact of permanent infrastructure from that of derived development for a more focused and efficient effort toward pollution reduction.
\end{abstract}

}

\maketitle

\IEEEdisplaynontitleabstractindextext

%
\IEEEpeerreviewmaketitle

\IEEEraisesectionheading{\section{Introduction}\label{sec:introduction}}

%
%
%
%
\IEEEPARstart{S}{tudies} in geography have changed their paradigms from descriptive to quantitative through the 'quantitative revolution' in the 1950s\cite{thedictionaryofhumangeography}. As pioneers in that revolution, Garrison and Marble\cite{Garrison1962THESO} developed a regression model in which transport network indices, for example the Betti number, the Alpha and Gamma numbers, and the diameter, etc., are described as linear combinations of five physical variables such as technological development, demographic level, size, shape, and relief. Relief is an ad hoc measure for the part in which the surface route distance is greater than the corresponding airline length. As a result, the structure of the transport network denoted as graphs with nodes and edges can be numerically calculated. 

Bodino\cite{Bodino1962} also discussed transport networks from the perspective of graphs; however, his discussion stemmed from the objective to minimise transportation cost, leading to the study of traffic flows, embodied as the Ford and Fulkerson theorem or the Gale theorem. Kansky\cite{kansky1963structure} systematically developed the measures of transport network structures in the Ph.D. thesis, where  
graph theory\cite{berge1982theory} was applied as a subfield of mathematics. The measures can be classified as non-ratio (Betti number, diameter), ratio (Alpha number, Gamma number) and individual elements (degree of connectivity, circuity).  

Starting from the 1970s, the transport network is no longer studied only as a static object, but as a growth mode subject to continuous development. Radke\cite{radke1977} used multiplier and critical distance to generate a stochastic model for network growth in 1977, which was then tested on the measures developed by Kansky, Garrison and Marble. Transportation was also actively discussed as a topic in social planning, especially in the 1980s when suburban centres exploded with urban development\cite{giuliano2017geography}, followed by changes in traffic flows\cite{newell1980} and travel patterns\cite{alma997397901811}. Meanwhile, spatial expansion required a large-scale view\cite{Xie_Levinson_2007} of the transport network, and new methods for increased complexity in randomness, heterogeneity, and modularity\cite{Solé2004}. The further development of graph theory\cite{chungspectral} has helped meet those requirements. 

For any random graph, in addition to the most regular k-complete and the most random Erdos-Renyi\cite{Bollobás1979} models, there are two typical graphs, neither too regular nor too random, called small-world and scale-free graphs, which were developed at the end of the 1990s. The small-world\cite{Watts_Strogatz_1998} describes an intermediate state where the characteristic path length decreases non-linearly as shortcuts while the clustering coefficient stays high, resulting in a sparse in space but clustered in connection small-world; the scale-free\cite{Barabási_Albert_1999} indicates a graph with the power-law distribution of node degrees. Newman\cite{doi:10.1137/S003614450342480} explained these theories and applications in detail in his review paper published in 2003.  Recent research that can be involved in the study of transport networks is more specific\cite{black2003transportation}. They are either building a framework, such as transit design\cite{Derrible_Kennedy_2011}, sprawl\cite{Barrington-Leigh_Millard-Ball_2015} or compact\cite{Stevenson_Thompson_deSá_Ewing_Mohan_McClure_Roberts_Tiwari_Giles-Corti_Sun_etal._2016} urban forms, or dealing with a concept, such as cities\cite{batty2013new}. 

Air pollution, which has been speculated as a health problem since ancient times, continues to be a burden on public health during the first two decades of this century\cite{Fowler_Brimblecombe_Burrows_Heal_Grennfelt_Stevenson_Jowett_Nemitz_Coyle_Liu_etal._2020}. By comparing the Beijing haze in 2012 with the London smog in the 1950s, air pollution changes significantly with time and location due to economic growth\cite{Jiang_Kim_Woo_2020}. Transport as a major source of air pollution has long been studied in vehicles and fuel consumption, which is a bottom-up approach. Fuel production and vehicle manufacturing\cite{Colvile_Hutchinson_Mindell_Warren_2001} are rarely included. The transport of pollutants between locations\cite{Chen_Chen_Zhu_Wang_Xie_2023}, the spillover effect\cite{Zeng_Chen_Dong_Liu_2023}, or the meteorological impact, such as wind (speed and direction), temperature, radiation, or precipitation\cite{Wijnands_Nice_Seneviratne_Thompson_Stevenson_2022}, are also studied.

Although the association between sprawl index and carbon dioxide emissions has been examined using automobile ownership\cite{Zhang_Chen_Li_Liu_2023}, top-down methods are not enough from the perspective of the association of transport networks with air pollution. The impact that facilitates or limits vehicle mobility remains understudied. The impact, though indirect, is more fundamental. If a strong relevance can be detected between static infrastructure and dynamic pollution, derived efforts through vehicle control could become more difficult when the network where vehicles reside remains the same. Therefore, the investigation of transport networks related to air pollution is significant.
\begin{figure*}[htbp]
 \centering \includegraphics[width=0.6\textwidth,keepaspectratio]{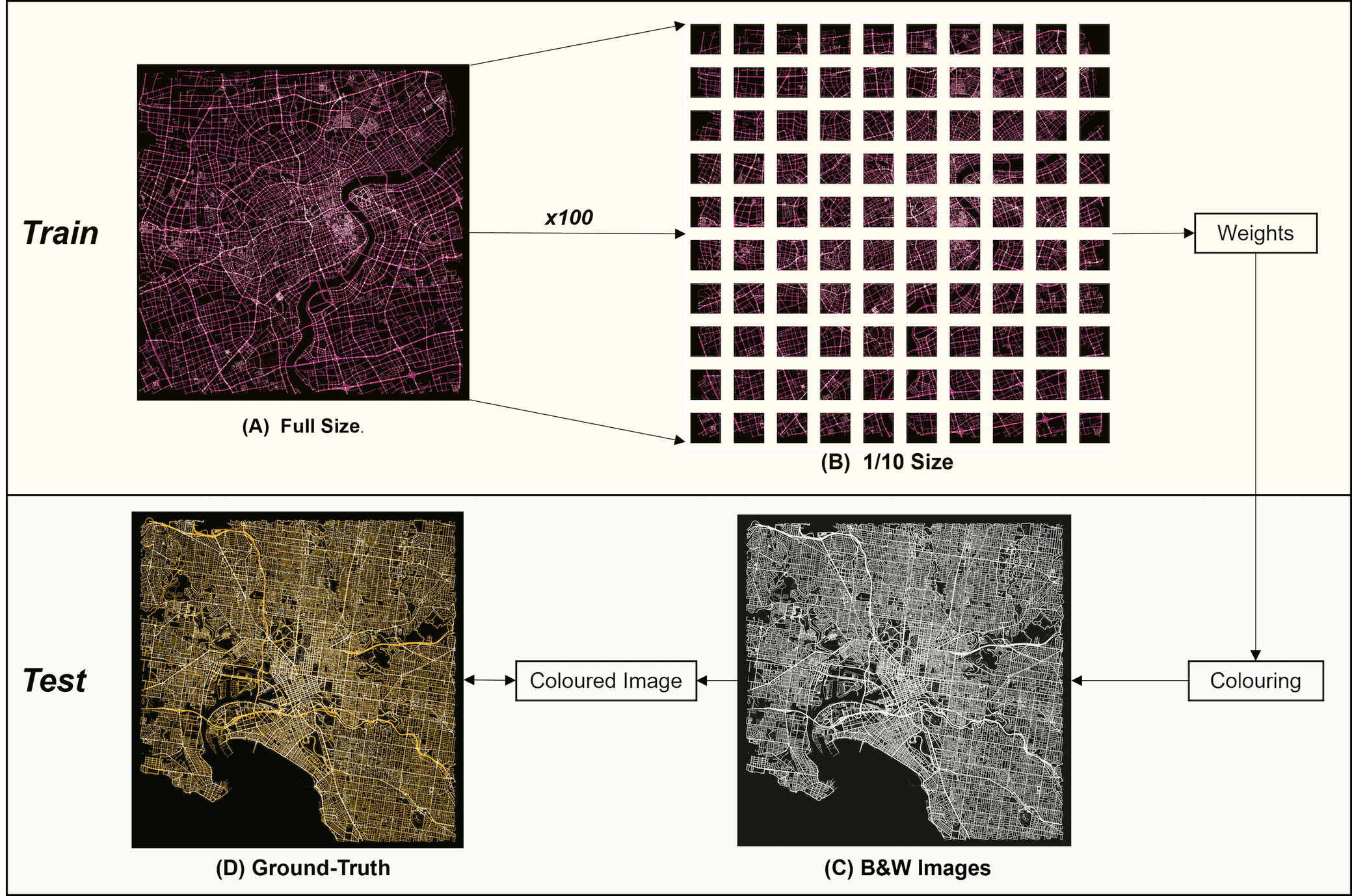}
  \caption{\textbf{Pollution-indicated transport network image interpretation.} The training process is divided into two stages: one uses full-size images sampled as image \textbf{A} to learn the overall properties, and the other uses $\frac{1}{10}$-size images as in \textbf{B}, which are generated from \textbf{A}, to learn the details. The weights learnt from the training are used to test B\&W images sampled as \textbf{C} in full and partial sizes. The resultant image will be compared to its ground truth as \textbf{D}.}
  \label{fig:fig-emis-p1}
\end{figure*}

Machine learning tools\cite{SAMAD2023119987} are widely used in recent studies of air pollution, such as random forest, XGboost, but computer vision and deep learning methods are used limitedly. In this study, a conditional generative adversarial network (cGAN) is used to study geometric patterns of transport networks in association with pollution through 0.3 million images, complemented by the topological characteristics analysed based on graph theory. Consequently, 12 indices on the transport network of $\sim$1700 cities are investigated for the overall association between the network and pollution in terms of geometry and topology. 7 out of 12 indices are revisited from the studies by Kansky, Garrison, and Marble; 5 are created through inspirations from the investigated geometric patterns or the small-world and scale-free models.


 



\section{Geometric Patterns: Pollution-Indicated Transport Network}
Transport networks can be viewed as geometric patterns in which shapes, orientations, and formed angles between line segments are concerned, or topological patterns in the way nodes are connected. Previous studies can only classify transport networks as typical geometric types, such as lattices or trees. Deep neural networks enable more detailed pattern recognition so that the transport network is parameterised as individually distinct types based on their geometric features, instead of any typical ones in a limited group of categories. 

The idea is to train a deep learning model specialised in image interpretation, Pix2Pix cGAN\cite{pix2pix2017}, to conduct the colouring\cite{colorizeWallner} task\cite{colorizeGuilbert} on roads according to the predicted concentrations of air pollution. The model is trained with images of transport networks with labelled concentrations. Black and white (B\&W) binary images are colourised on the lines of transport networks of 1692\footnote{1689 for NO$_2$.} global cities based on the street-level concentrations of inhalable particulate matter PM$_{2.5}$ and nitrogen dioxide NO$_{2}$. The model first converts the colour images from $RGB$ to $Lab$ space to separate air pollution from the transport network as \Cref{spaceconversion}.
\begin{equation}
\label{spaceconversion}
    RGB\rightarrow{L+a\&b}
\end{equation}
The $L$ channel has only intensity that represents the geometric structure of transport networks; while the $a\&b$ channels have colours that indicate different levels of pollution. The training process enables the model to recognise the relationship between the $L$ channel and the $a\&b$ channels; or the relationship between geometric transport networks and air pollution. The trained weights are used to predict $a\&b$ channels of the binary images of transport networks based on their $L$ channels. In this way, the pattern of geometric network-induced pollution can be identified. As in Fig.~\ref{fig:fig-emis-p1}, the upper half indicates the training process, while the lower half is the test that colourizes the B\&W binary images using trained weights and compares them to the intended colour images. 

\subsection{Data}
The colour images used in this investigation are prepared by integrating raster pollution data PM$_{2.5}$\cite{pm25data2019} \& NO$_2$\cite{cooper_2022_5484305} with vector maps obtained from OpenStreetMap\cite{OpenStreetMap} using a general algorithm specially developed by the author for efficient data fusion\cite{10874104}. Pollution data are annual averages for 2019 to avoid the pandemic. As in Table~\ref{tab-trainandtest}, the data contains PM$_{2.5}$ and NO$_2$ in 1692 and 1689 global cities, respectively, in which pollution concentrations are indicated in lines with the colour palette of reversed plasma and road types with line widths. The width values are taken from the OSMnx\cite{osmnxBoeing} tutorial to display the visible difference between the roads to walk and drive.

\subsection{Train \& Test}
cGAN trains full-size images in the first stage epochs to capture general features by compressing the images to fit the encoder of the cGAN architecture, where fine features are lost. However, image size is critical for the performance of deep learning models, given the same resolution\cite{10410967}. Therefore, after the first-stage epochs, every image is divided into 100 smaller images to continue the training until the loss curves stabilise to plateaus again. In the test stage, the weights learnt from training are used to infer images in full and smaller sizes in 110 cities. The details of the size and number of images for training and testing are listed in Table~\ref{tab-trainandtest}. Binary images are used purposely in the test instead of the $L$ channels to exclude any residual information about pollution that could be embedded in the greyscale images. Some trial results have indicated that the $L$ channels facilitate the achievement of high accuracy in prediction due to residual information. Since we intend to obtain the standalone effect of geometric patterns, this facilitation is prevented by inferring binary images.

\begin{table}[htbp]
\caption{\textbf{Data used in image interpretation using cGAN.} 1692 cities are the full-range dataset. NO$_2$ has one city point missing in the original data, and two missing on transport network images. The number of the full-size and $\frac{1}{10}$-size images used in training and testing are listed in the table.}
\begin{center}
\begin{tabular}{lcc}
\hline
 & \textbf{PM$_{2.5}$} & \textbf{NO$_2$}  \\
\hline
Number of Cities & 1692 & 1689 \\
Data Type & \multicolumn{2}{c}{color \& binary images}  \\
Line Colour & \multicolumn{2}{c}{reversed plasma\cite{480803}}  \\
Line Width & \multicolumn{2}{c}{\makecell{0.5: footway,steps,pedestrian,path,track\\2: service; 3: residential; 5: primary; 6: motorway}} \\
\hline
\multicolumn{3}{c}{\textbf{Train}}\\
\hline
Number of Cities & 1582 & 1579 \\
\multicolumn{3}{c}{\textit{1$^{st}$ Stage}} \\
Image Size$_1$ (S1) & 10240x10240 & 10240x10240 \\
Number of S1 & 1582 & 1579 \\
\multicolumn{3}{c}{\textit{2$^{nd}$ Stage}}\\
Image Size$_2$ (S2) & 1024x1024 & 1024x1024 \\
Number of S2 & 158200 & 157900 \\
\hline
\multicolumn{3}{c}{\textbf{Test}}\\
\hline
Number of Cities & 110 & 110 \\
S1 & 10240x10240 & 10240x10240 \\
Number of S1 & 110 & 110 \\
S2 & 1024x1024 & 1024x1024 \\
Number of S2 & 11000 & 11000 \\
\hline
\end{tabular}
\label{tab-trainandtest}
\end{center}
\end{table}

\begin{figure}[htbp]
  \centering 
  \includegraphics[width=0.4\textwidth,keepaspectratio]{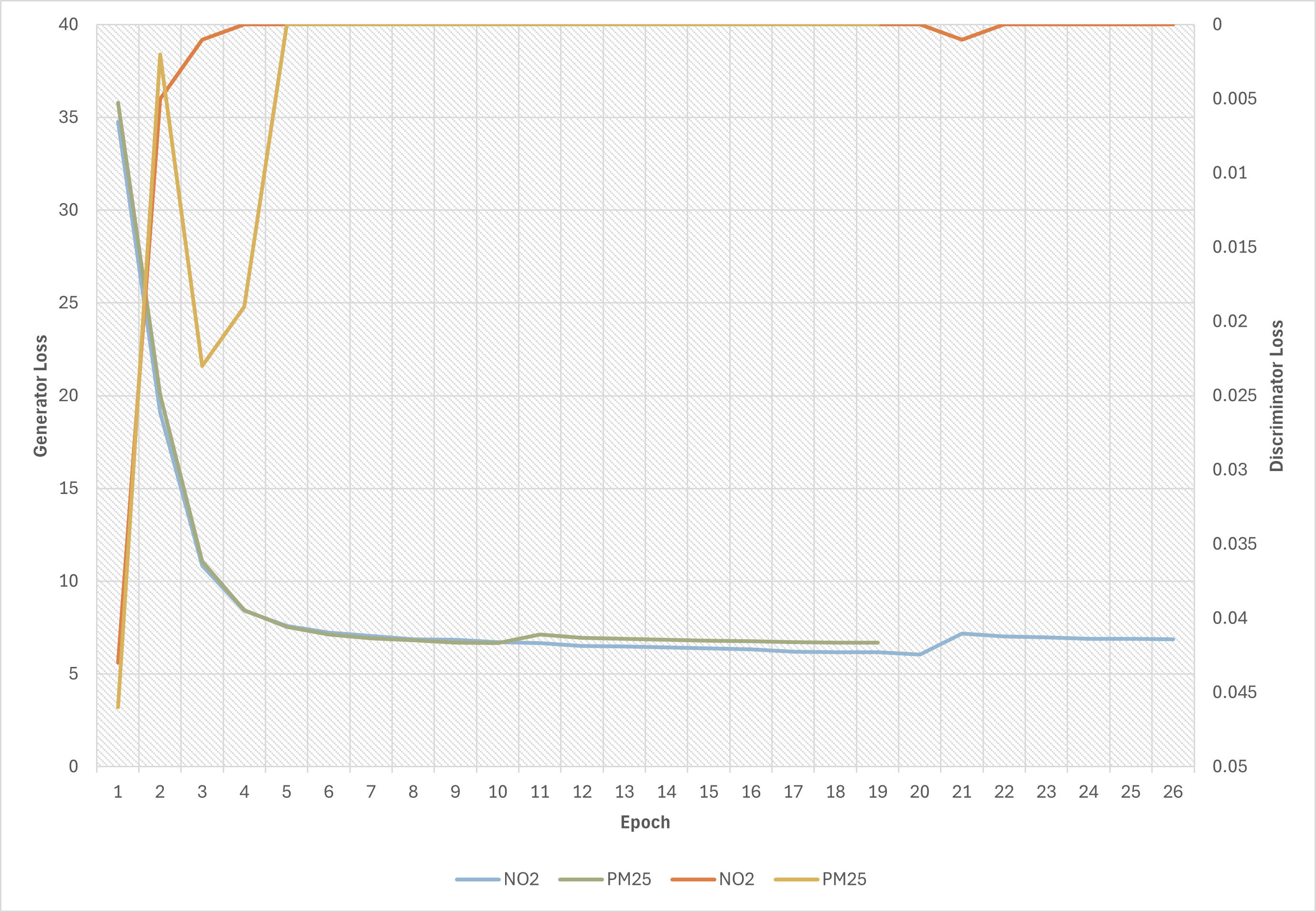}
  \caption{\textbf{Loss curves from the training process of image interpretation.} cGAN trains one generator and one discriminator at the same time, the loss curves for the generator are aligned with the left coordinate; that for the discriminator are aligned with the right one. The generator loss curves for PM$_{2.5}$ and NO$_2$ are in green and blue respectively. The discriminator loss curves for PM$_{2.5}$ and NO$_2$ are in yellow and red respectively. The aligned "bumps" shown on two loss curves for NO$_2$ indicate the change of training stages from the full-size images to the $\frac{1}{10}$-size. The bump on the generator loss curve for PM$_{2.5}$ indicates the same change, however its corresponding bump is invisible on the discriminator loss curve for PM$_{2.5}$.}
  \label{fig:fig-emis-p2}
\end{figure}

\subsection{Result}
\begin{figure*}[htbp]
  \centering \includegraphics[width=0.6\textwidth,keepaspectratio]{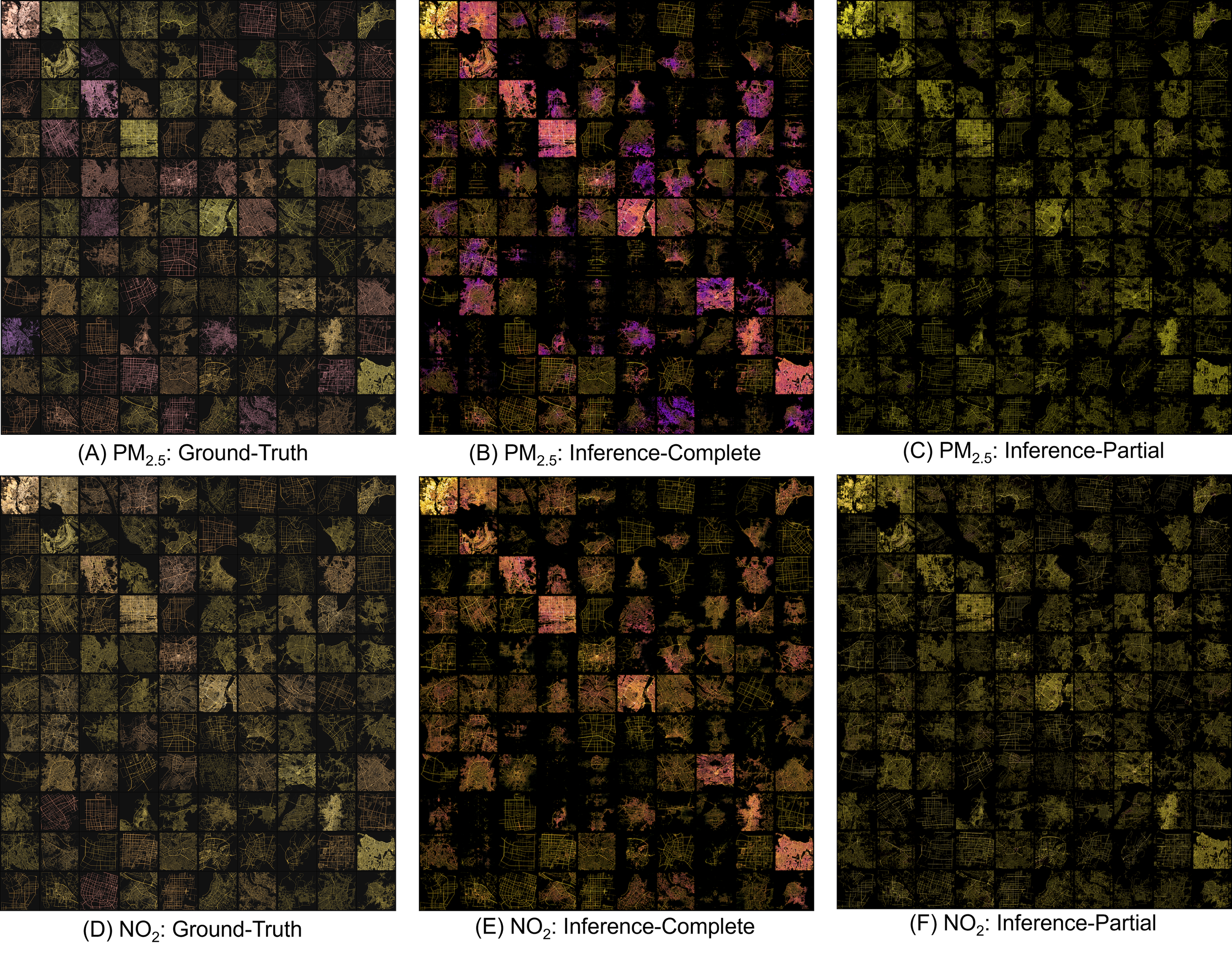}
  \caption{\textbf{Results of inferences on 110 cities images for PM$_{2.5}$ and NO$_2$.} Ground-truth images for 110 cities are listed in the 1$^{st}$ column where \textbf{A} is for PM$_{2.5}$ and \textbf{D} for NO$_2$. Inferences on full-size images are listed in the 2$^{nd}$ column where \textbf{B} is for PM$_{2.5}$ and \textbf{E} for NO$_2$. Inferences on $\frac{1}{10}$-size images and then combined to form the full size are listed in the 3$^{rd}$ column where \textbf{C} is for PM$_{2.5}$ and \textbf{F} for NO$_2$. Similar patterns can be found between \textbf{B} and \textbf{E}. Comparison of patterns between \textbf{C} and \textbf{F} requires enlarged views which are included in Supplementary \Crefrange{DEPM}{CNNO}.}
  \label{fig:fig-emis-p3}
\end{figure*}
As the two-stage training was applied, the loss curves of the generator and discriminator are plotted by combining the curves of two stages. As in Fig.~\ref{fig:fig-emis-p2}, cGAN runs 19 epochs in PM$_{2.5}$, in which 11 epochs are in the first stage, and 26 in NO$_2$, with 21 in the first stage. The small "bumps" aligned with the two curves of NO$_2$ indicate the shift from the first stage to the second. Training stops at a similar loss for both stages of the curves.

The weights from the last training epoch are used to infer the test images of 110 cities at full size and $\frac{1}{10}$ size as in the middle and third columns, respectively, in Fig.~\ref{fig:fig-emis-p3}. Although the ground truth images are visibly different for PM$_{2.5}$ and NO$_2$, similar patterns can be identified in the resultant images in full size, indicating that pollution induced solely by the effect of the geometry of the transport networks does exist. More detailed observations can be made as follows in the combined images of the size $\frac{1}{10}$ by individually amplifying the images in the third column.

\begin{figure*}[htbp]
\centering
\begin{minipage}[t]{0.2\textwidth}
    \begin{tikzpicture}[spy using outlines={circle,white,magnification=5,size=3cm, connect spies}]
    \node{\includegraphics[width=\textwidth]{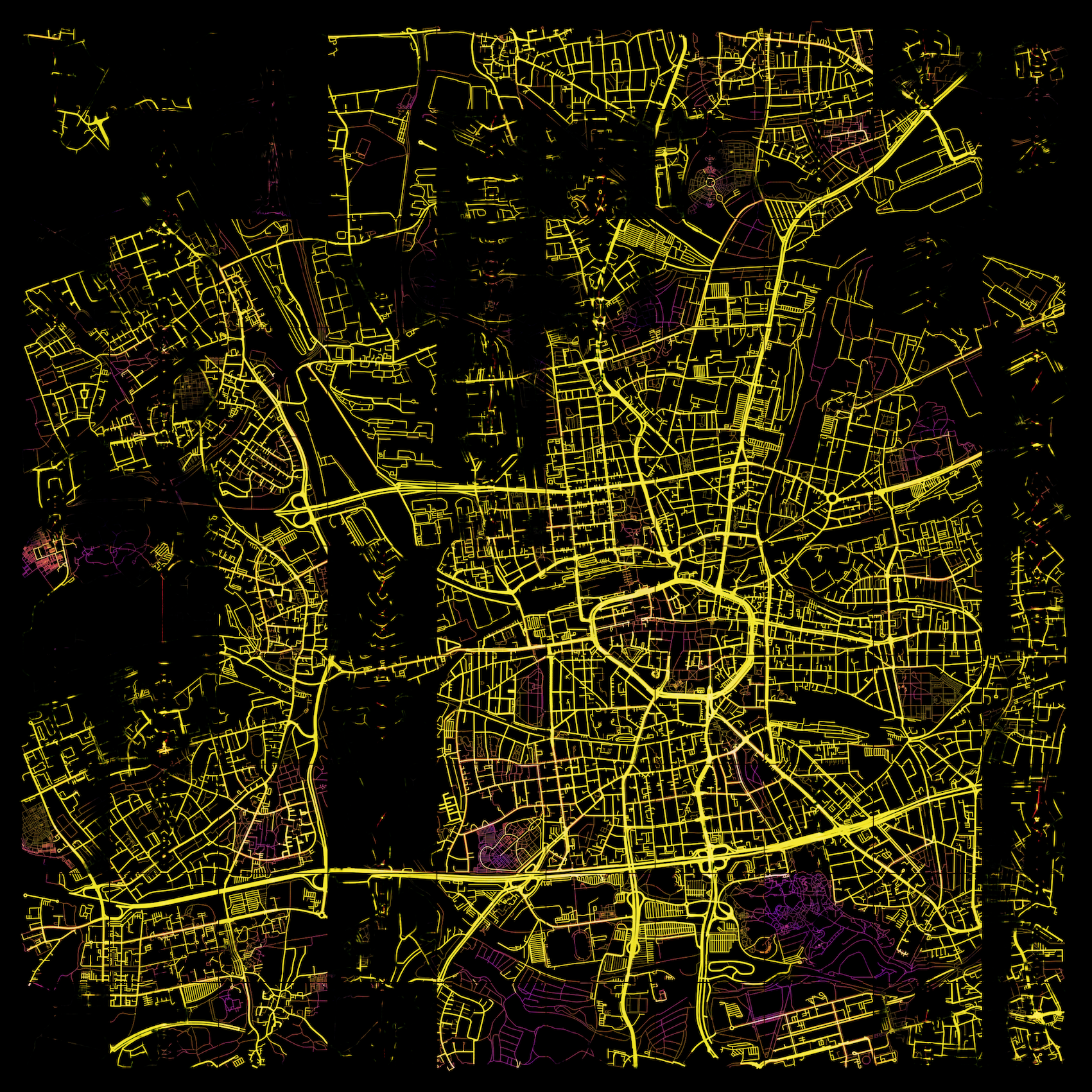}};
    \end{tikzpicture}
    \caption{\newline PM$_{2.5}$:Dortmund/DE. 
    }
    \label{DEPM}
\end{minipage}
\begin{minipage}[t]{0.2\textwidth}
    \begin{tikzpicture}[spy using outlines={circle,white,magnification=5,size=3cm, connect spies}]
    \node{\includegraphics[width=\textwidth]{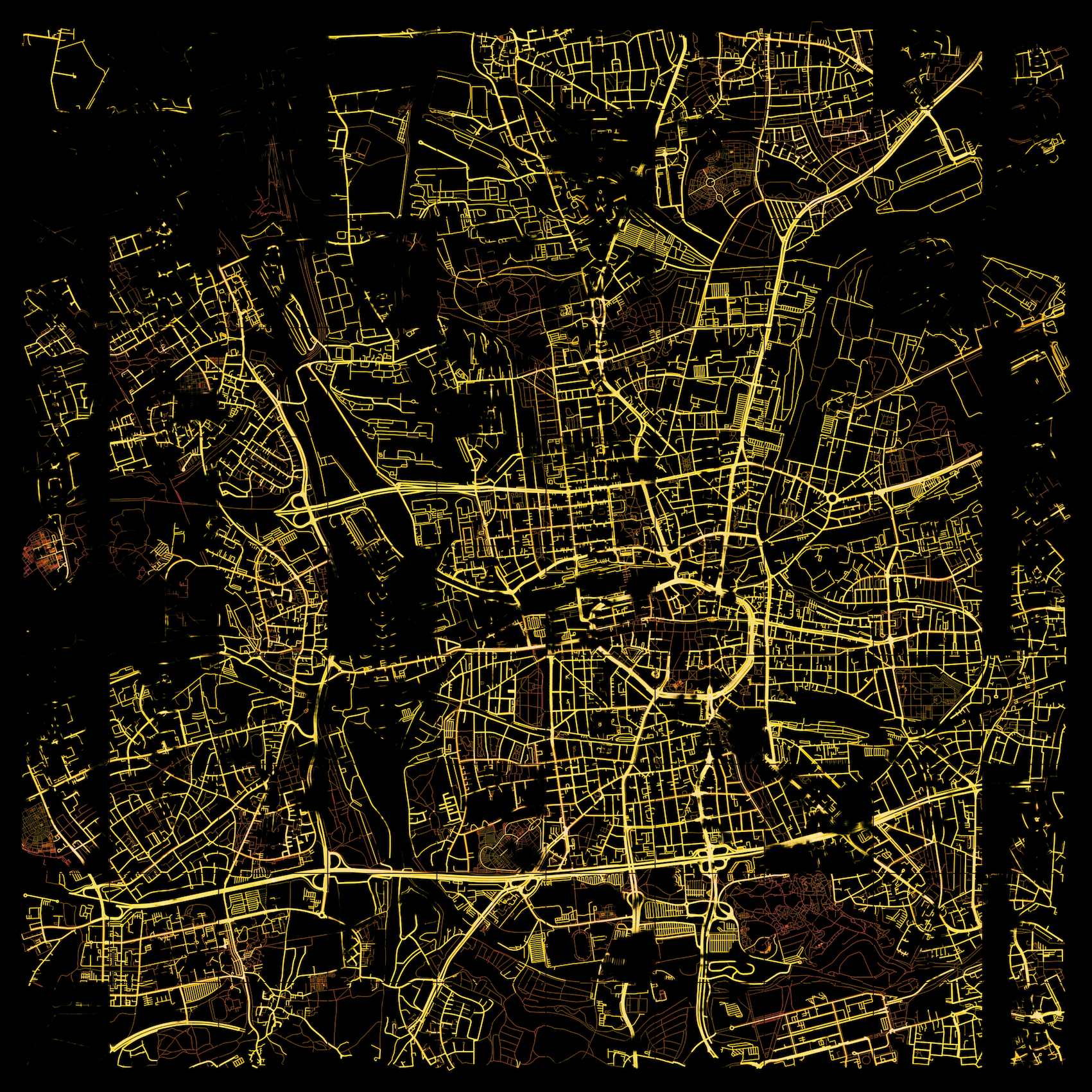}};
    \end{tikzpicture}
    \caption{\newline NO$_{2}$:Dortmund/DE. 
    }
    \label{DENO}
\end{minipage}
\begin{minipage}[t]{0.2\textwidth}
    \begin{tikzpicture}[spy using outlines={circle,white,magnification=5,size=3cm, connect spies}]
    \node {\includegraphics[width=\textwidth]{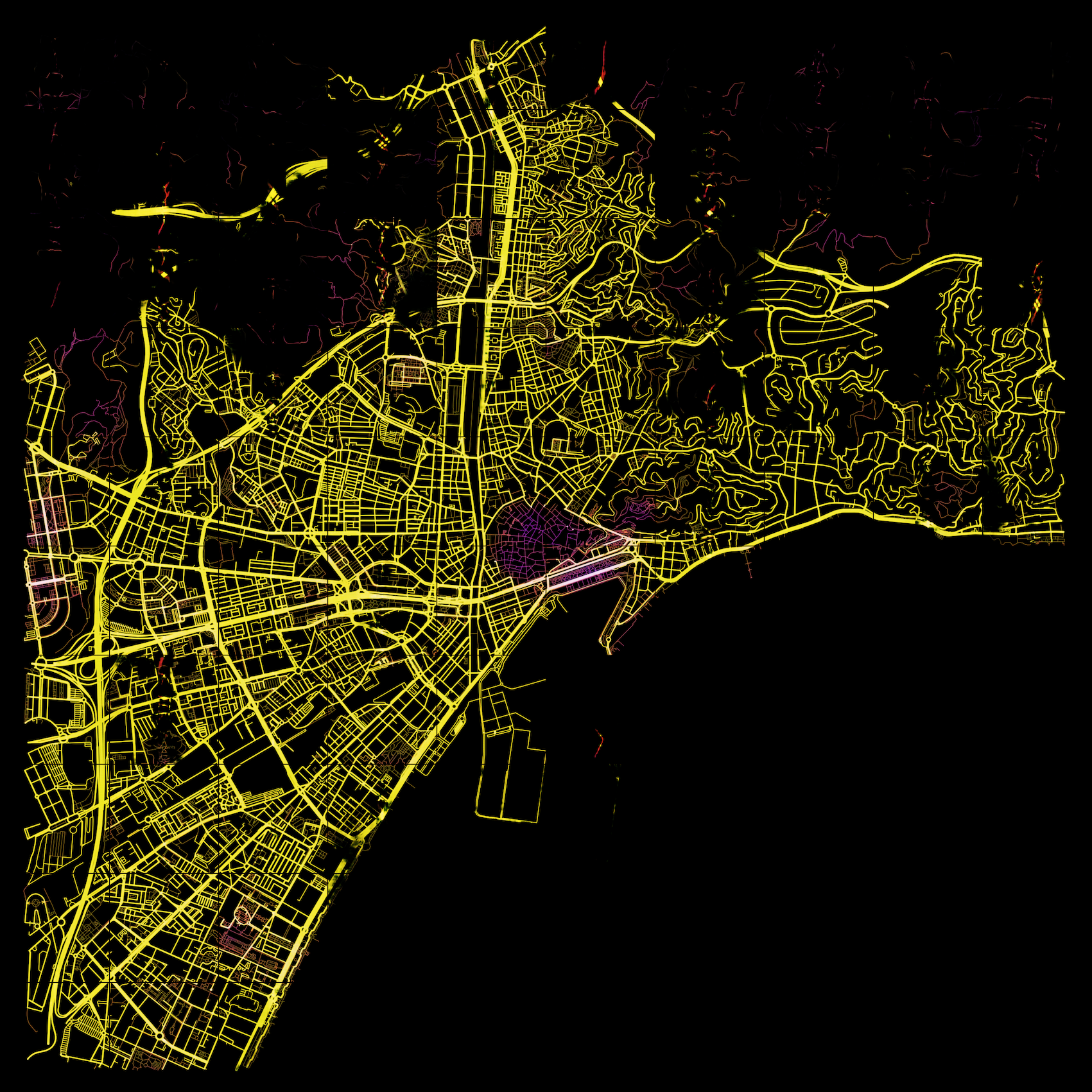}};
    \end{tikzpicture}
    \footnotesize
    \caption{\newline PM$_{2.5}$:Malaga/ES 
    }
    \label{ESPM}
\end{minipage}
\begin{minipage}[t]{0.2\textwidth}
    \begin{tikzpicture}[spy using outlines={circle,white,magnification=5,size=3cm, connect spies}]
    \node {\includegraphics[width=\textwidth]{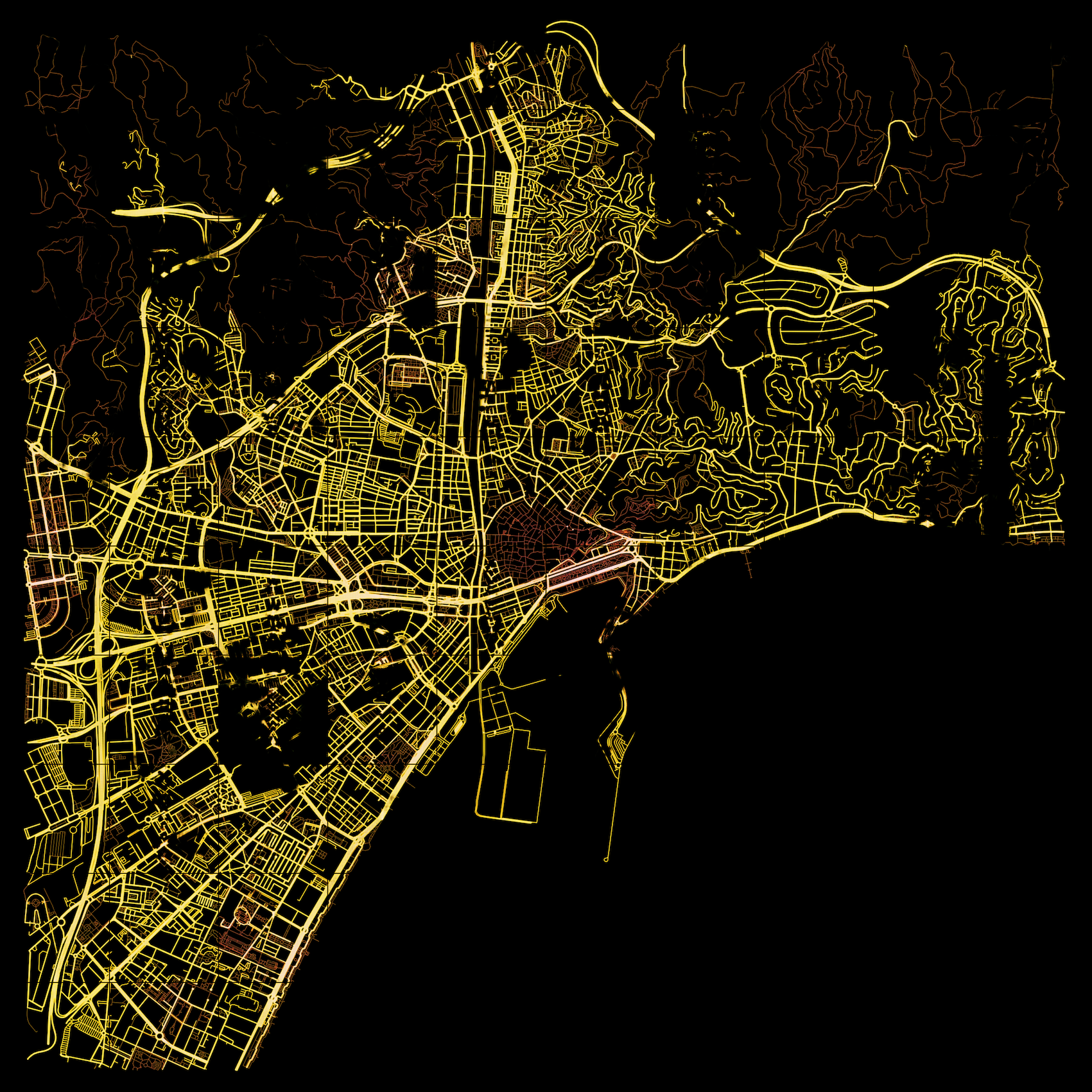}};
    \end{tikzpicture}
    \caption{\newline NO$_{2}$:Malaga/ES.
    }
    \label{ESNO}
\end{minipage}
\end{figure*}

1. Roads for walking, for example, footpath, steps, pedestrians, etc., tend to show higher concentrations of pollution compared to drivable roads, for example, service, residential, etc., especially near the sudden shift of road types (Dortmund as Fig.~\ref{DEPM}\text{--}\ref{DENO}; Malaga as Fig.~\ref{ESPM}\text{--}\ref{ESNO}).

\begin{figure*}
\centering
\begin{minipage}[t]{0.2\textwidth}
    \begin{tikzpicture}[spy using outlines={circle,white,magnification=5,size=3cm, connect spies}]
    \node{\includegraphics[width=\textwidth]{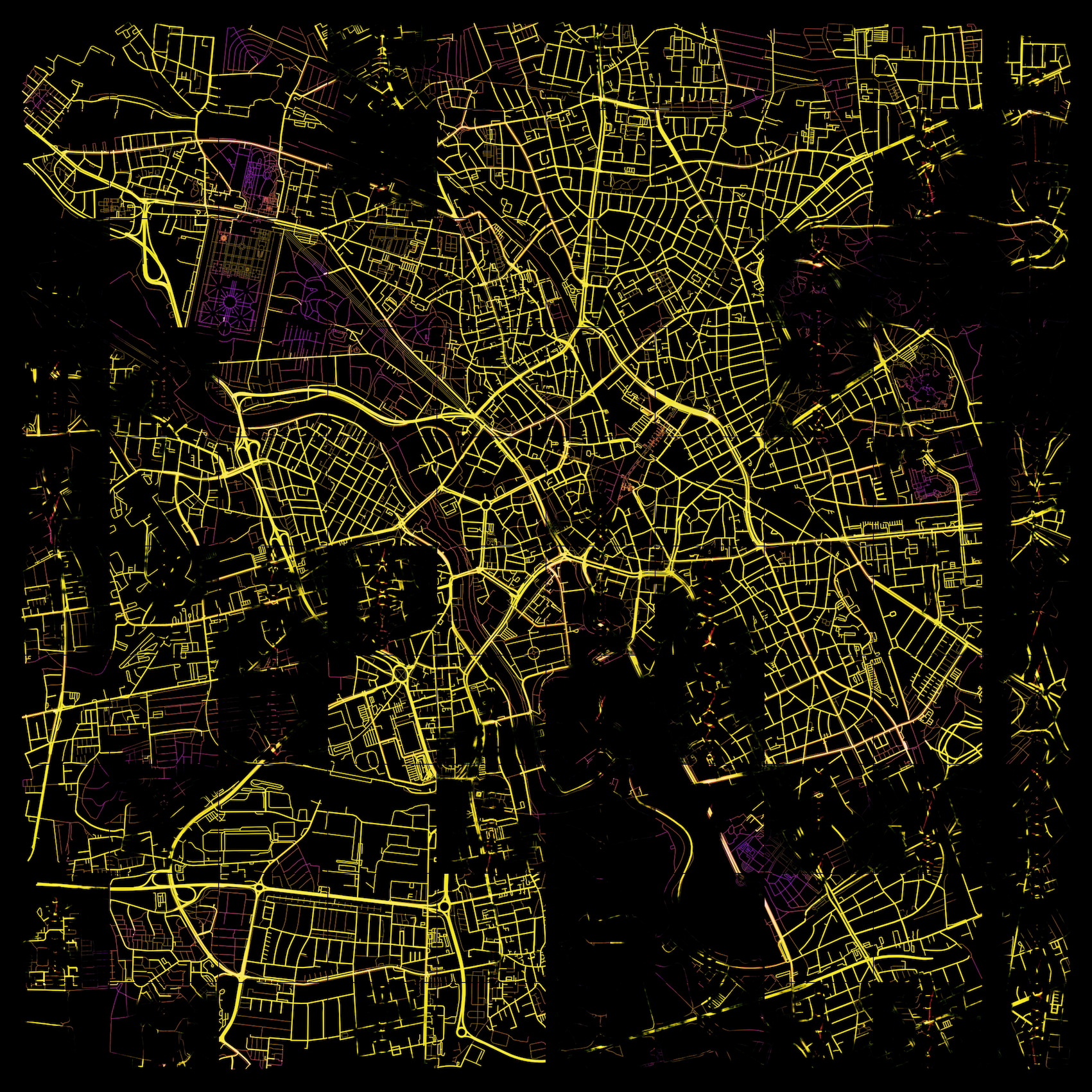}};
    \end{tikzpicture}
    \caption{\newline{PM$_{2.5}$:Djibouti/DJI.} 
    }
    \label{DJIPM}
\end{minipage}
\begin{minipage}[t]{0.2\textwidth}
    \begin{tikzpicture}[spy using outlines={circle,white,magnification=5,size=3cm, connect spies}]
    \node {\includegraphics[width=\textwidth]{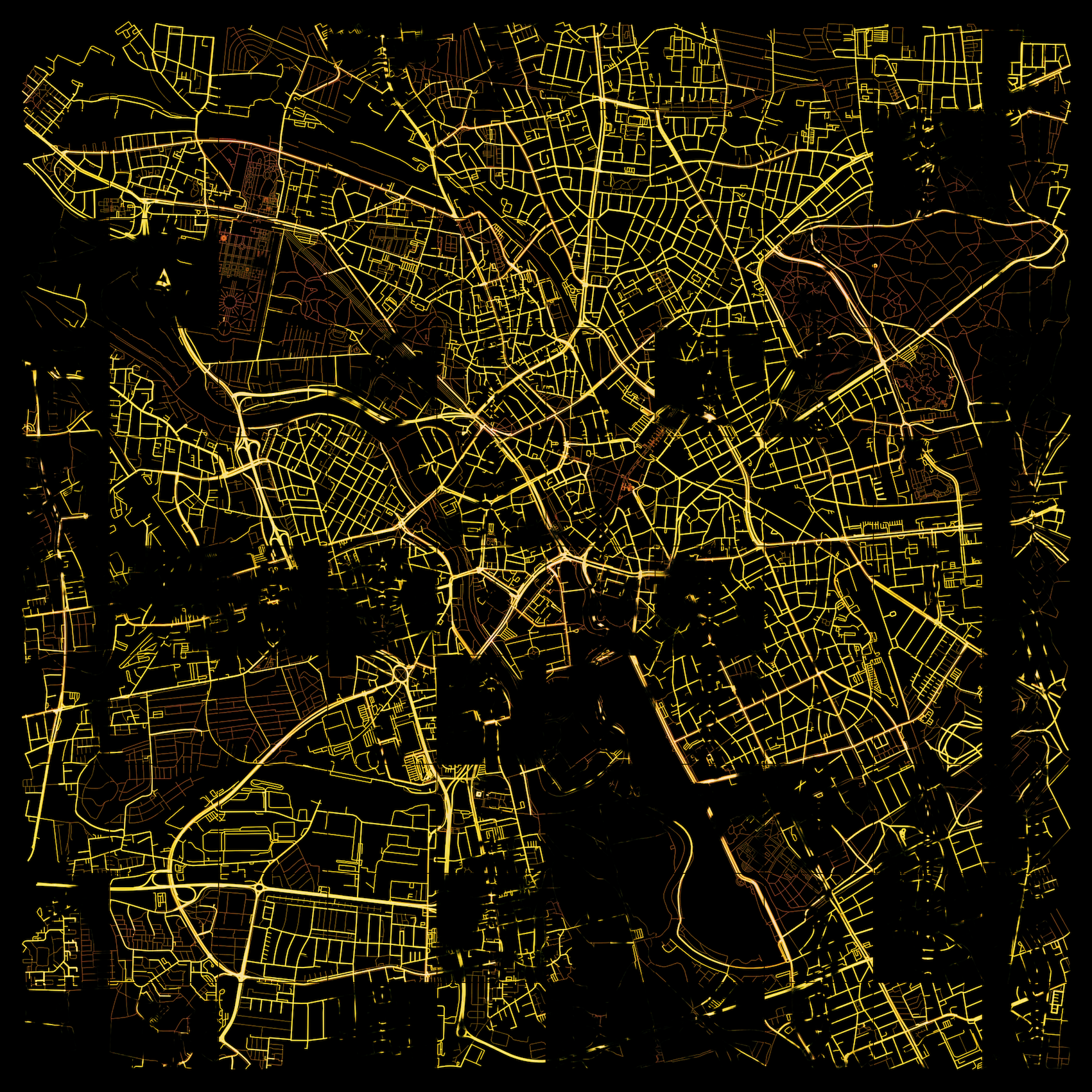}};
    \end{tikzpicture}
    \caption{\newline{NO$_{2}$:Djibouti/DJI.} 
    }
    \label{DJINO}
\end{minipage}
\begin{minipage}[t]{0.2\textwidth}
    \begin{tikzpicture}[spy using outlines={circle,white,magnification=5,size=3cm, connect spies}]
    \node{\includegraphics[width=\textwidth]{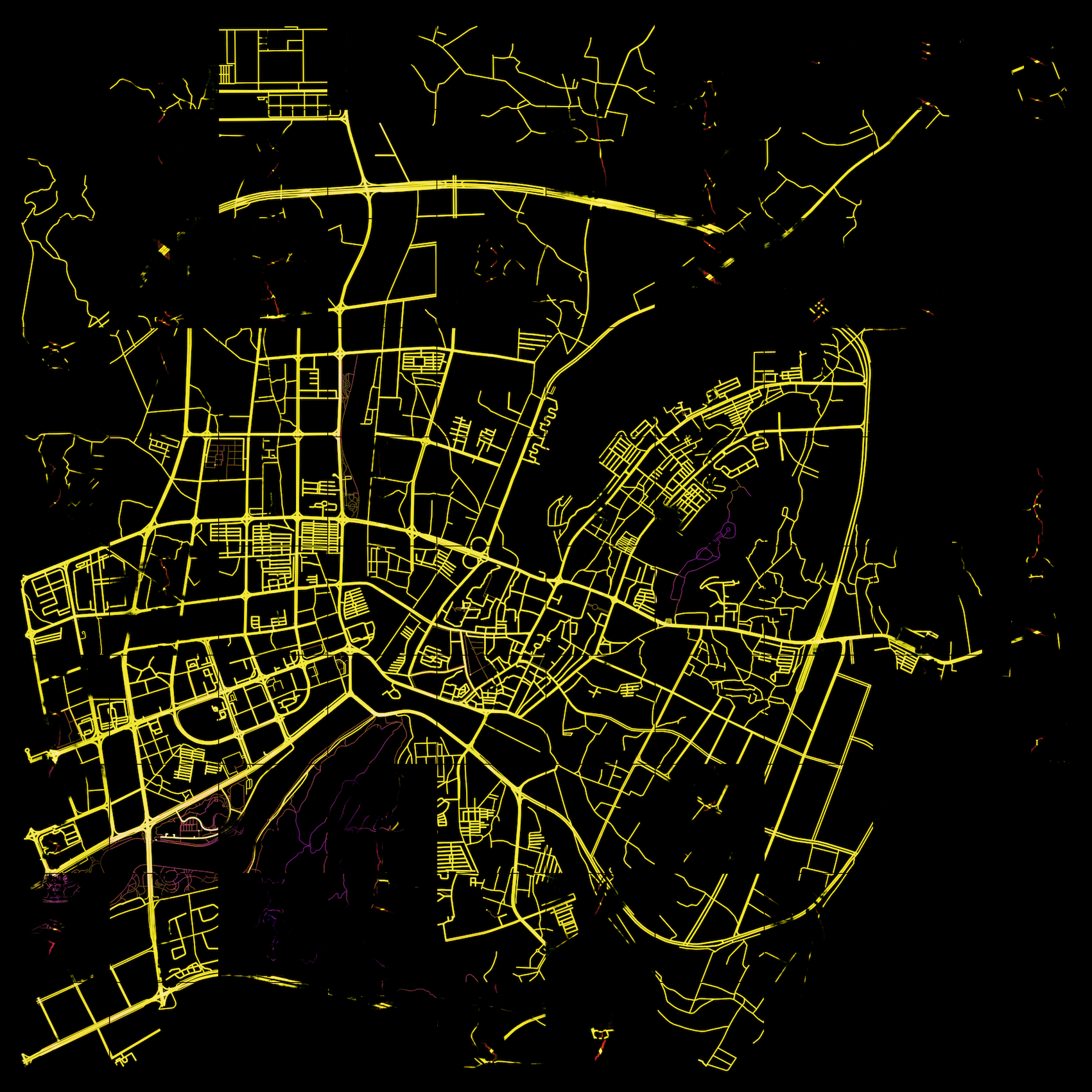}};
    \end{tikzpicture}
    \caption{\newline{PM$_{2.5}$:Hebi/CN.} 
    }
    \label{CNPM}
\end{minipage}
\begin{minipage}[t]{0.2\textwidth}
    \begin{tikzpicture}[spy using outlines={circle,white,magnification=5,size=3cm, connect spies}]
    \node {\includegraphics[width=\textwidth]{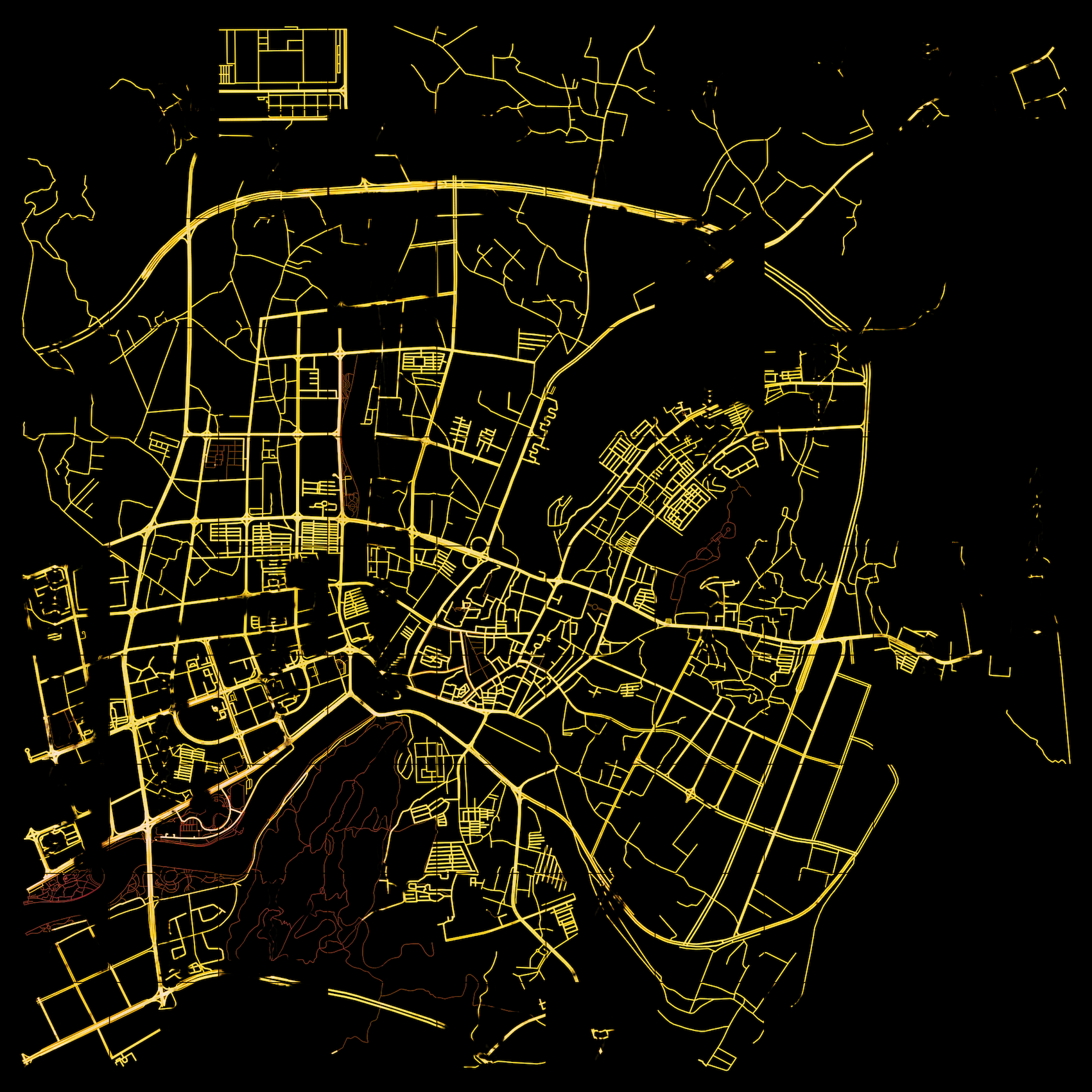}};
    \end{tikzpicture}
    \caption{\newline{NO$_{2}$:Hebi/CN.} 
    }
    \label{CNNO}
\end{minipage}
\end{figure*}

2. The curvy roads tend to indicate higher concentrations compared to straight roads (Djibouti as Fig.~\ref{DJIPM}\text{--}\ref{DJINO}; Hebi as Fig.~\ref{CNPM}\text{--}\ref{CNNO}), and disordered roads that bring a higher probability of randomness to the network structure tend to show higher concentrations (Malaga as Fig.~\ref{ESPM}\text{--}\ref{ESNO}; Djibouti as Fig.~\ref{DJIPM}\text{--}\ref{DJINO}).


\newcommand\AddLabel[1]{%
  \refstepcounter{equation}
  (\theequation)
  \label{#1}
}
\newcolumntype{M}{>{$\displaystyle}c<{$}}
\newcolumntype{L}{>{\collectcell\AddLabel}r<{\endcollectcell}}
\begin{table*} 
	\centering
	\caption{\textbf{Functions to describe transport networks in 12 indices.} \Crefrange{eq-1}{eq-7} corresponding to IDX~$1\text{--}7$ are reused from the research in the 1970s. The source can be found in \cite{radke1977}. \Crefrange{eq-8}{eq-12} are created for this research, in which \Crefrange{eq-8}{eq-9-2} corresponding to IDX~{8\text{--}9} are designed to quantify two kinds of geometric patterns observed in image interpretation; \Crefrange{eq-10-1}{eq-11-3} corresponding to IDX~{10\text{--}11} are inspired by the small-world and scale-free models; \Cref{eq-12} for IDX~$12$ is to confirm the overall distribution of full-range cities in terms of connectivity. 
	}
	\label{SCIENCE-T1} 
	\begin{tabular}{llML} 
		\hline
		\textit{No.} & \textit{Index (IDX)} & \multicolumn{1}{l}{\textit{Function}} & \multicolumn{1}{l}{} \\
		\hline
		1 & Mean Local Degrees & deg_{mean}=\frac{2m}{n} & eq-1 \\
		2 & First Betti Number & \mu=m-n+p & eq-2 \\
		3 & Alpha Index & \alpha=\frac{2\mu}{(n-1)(n-2)} & eq-3 \\
        4 & Gamma Index & \gamma=\frac{2m}{n(n-1)} & eq-4 \\
        5 & Redundancy Ratio & RI=\frac{n^{2}}{SDI} & eq-5 \\
        6 & Diameter & diam(G)=max_{u,v \in G}(d_{geod}(u,v)) & eq-6 \\
        7 & System Dispersion Index & SDI=\sum\limits_{u=1}^{n}\sum\limits_{v=1}^{n}d_{geod}(u,v) & eq-7 \\
        8 & Ratio of the Number of Walk to Drive Paths & RTI=\frac{\sum\limits_{u,v\in W_{k}}D(u,v)}{\sum\limits_{u,v\in D_{r}}D(u,v)}, W_{k} \cap D_{r} =\emptyset & eq-8 \\
        9 & Ricci Curvature\cite{RicciCurve} & \kappa(x,y)=1-\frac{W(m_x,m_y)}{d(x,y)} & eq-9-1 \\
         & & \kappa={\arg\max}_{\kappa_{i}}P(\kappa_{E}={\kappa_i}), i\in[1,m] & eq-9-2 \\
        10 & Clustering Coefficient\cite{cluster} & c_u=\frac{2T_u(v)}{deg(v)(deg(v)-1)} & eq-10-1 \\
         & & c={\arg\max}_{c_{i}}P(c_{V}={c_i}), i \in [1,n] & eq-10-2 \\
        11 & Scale-free Index & n_k\propto k^{-\beta}, \beta>1 & eq-11-1 \\
         & & \log{n_k}\approx C -\beta\log{k}, C=\text{const.} & eq-11-2 \\
         & & SFI=\frac{{\arg\min}_{n_{i}}P(n_{k}={n_i})}{{\arg\max}_{n_{i}}P(n_{k}={n_i})} , i \in [0,t] & eq-11-3 \\
        12 & Graph Connectivity & num_{connect}=\left\{\begin{array}{rcl}
            1, & \mbox{if} & \text{weakly-connected}\\ 
            2, & \mbox{if} & \text{semi-connected}\\
            3, & \mbox{if} & \text{strongly-connected}\\
            4, & \mbox{if} & \text{bi-connected}\end{array}\right. & eq-12 \\
		\hline
	\end{tabular}
\end{table*}

\section{Pollution-indicated Transport Network in 12 Indices}
\begin{figure*}[htbp]
  \centering \includegraphics[width=0.6\textwidth,keepaspectratio]{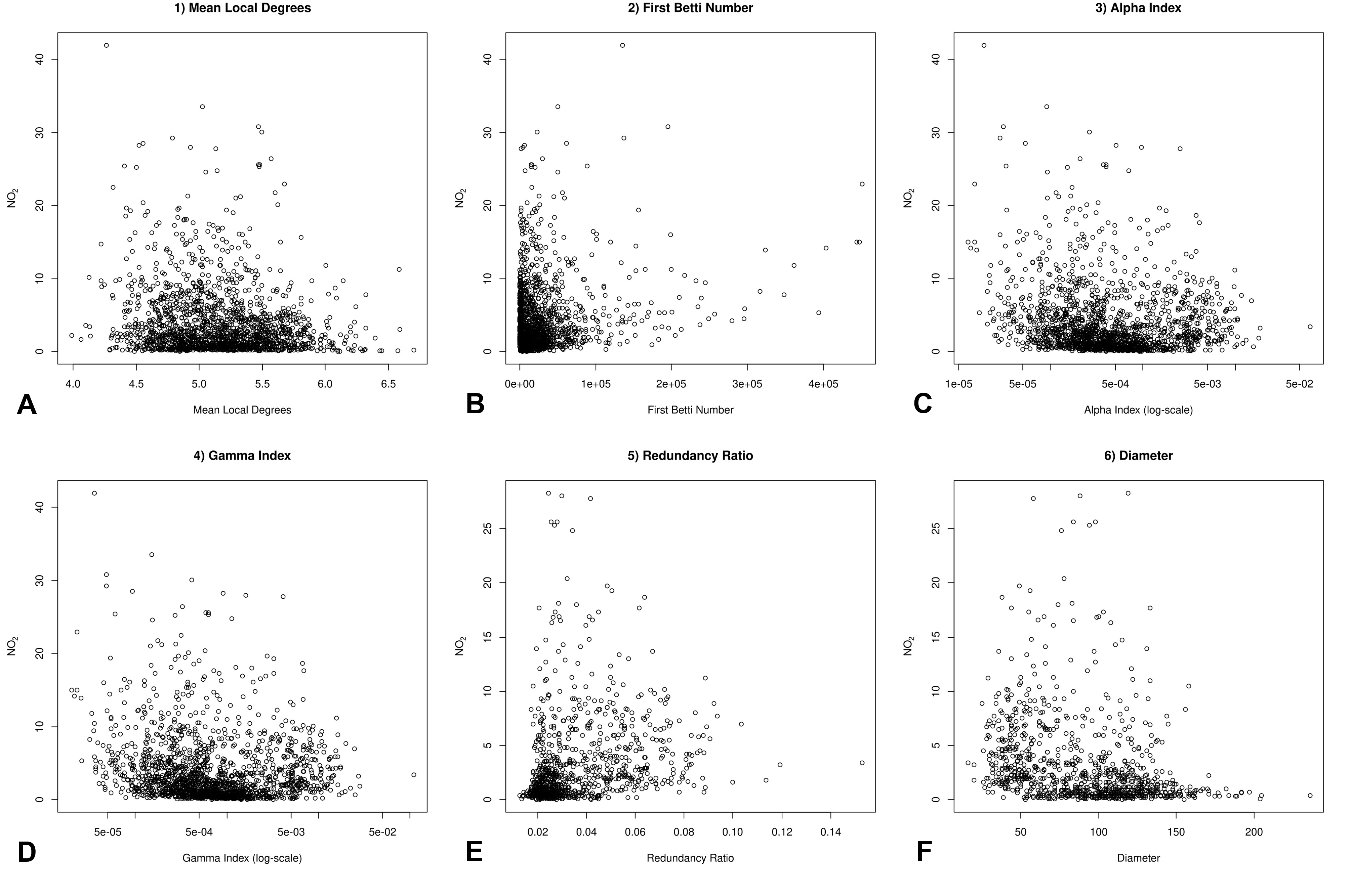}
  \caption{\textbf{Transport networks in 12 indices:~IDX~1\text{--}6.}  IDX~$1\text{--}4$ values of 1691 city points are plotted against NO$_2$ as in \textbf{A} to \textbf{D}, in which a logarithmic scale is used in \textbf{C} and \textbf{D} to accommodate one extreme point. 1691 cities are the full range data for NO$_2$ and 1689 cities are the full range image data for NO$_2$. IDX~$5\text{--}6$ values of 952 city points are plotted against NO$_2$ as in \textbf{E} to \textbf{F}. 952 cities are sampled from the full range to improve the efficiency compromised by the computing geodesic distances.}
  \label{fig:fig-emis-p1-p6}
\end{figure*}

\begin{figure*}[htbp]
  \centering \includegraphics[width=0.6\textwidth,keepaspectratio]{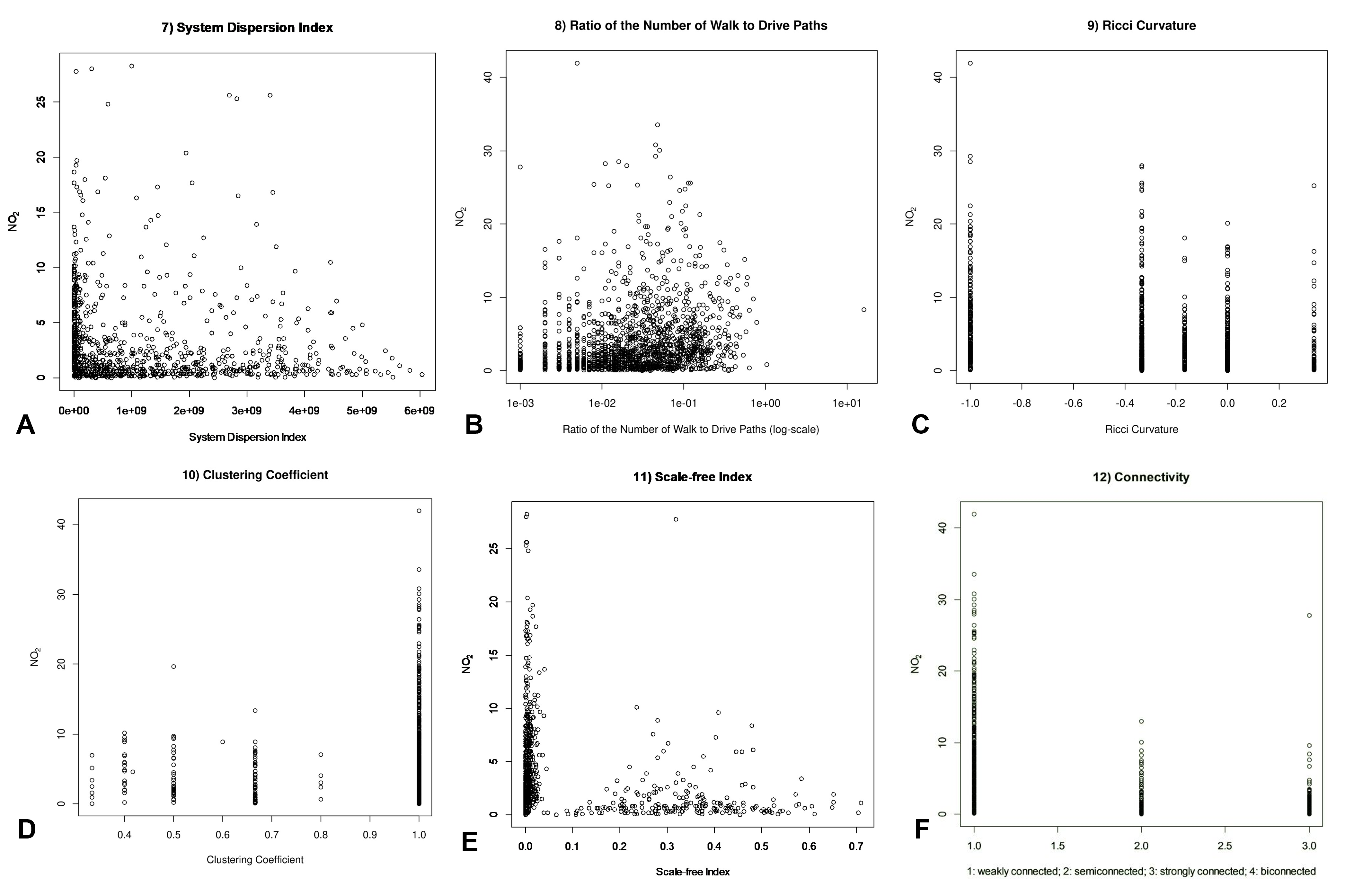}
  \caption{\textbf{Transport networks in 12 indices:~IDX~7\text{--}12.} IDX~$7$ and IDX~$11$ values of 952 city points are plotted against NO$_2$ as in \textbf{A} and \textbf{E}. IDX~$8$, IDX~$10$, and IDX~$12$ values of 1691 city points are plotted against NO$_2$ as in \textbf{B}, \textbf{D}, and \textbf{F}, in which a logarithmic scale is used in \textbf{B} to accommodate one extreme point. IDX~$9$ values of 1223 city points are plotted against NO$_2$ as in \textbf{C} to improve the efficiency compromised by the computing of curvatures.}
  \label{fig:fig-emis-p7-p12} 
\end{figure*}

We revisit the network parameters proposed by Garrison, Marble, and Kansky, summarized by Radke as 7 indices denoted as IDX~$1\text{--}7$, as \Crefrange{eq-1}{eq-7} in \Cref{SCIENCE-T1}, to explore new insights in association with air pollution. 5 indices, IDX~$8\text{--}12$, are developed as \Crefrange{eq-8}{eq-12} in an attempt to be inspired by the small-world and scale-free models, or the geometric patterns identified in the last section. The number of nodes (or "vertices") is defined as $n$, the edges as $m$, and the subgraph as $p=1$ for the networks within the scope of this research are all connected graphs. In any graph $G$, the geodesic distance between the nodes $u$ and $v$ is indicated as $d_{geod}(u,v)$. The values calculated from the sampled cities are plotted with respect to NO$_2$ as in Fig.~\ref{fig:fig-emis-p1-p6}\text{--}\ref{fig:fig-emis-p7-p12}.

IDX~$1\text{--}4$ represent an estimation for connectivity; the larger the index, the more connected the network. The logarithmic scale is used to fit one extreme case in the display of IDX~$3\text{--}4$ in Fig.~\ref{fig:fig-emis-p1-p6}: a $183$~km$^2$ county-level Bokaro steel city in India that has a small value of $n$, resulting in outliers of the two highly correlated indices. The mean local degree (IDX~$1$) for most cities falls in the range of $(4.5, 6)$. The first Betti number (IDX~$2$), or cyclomatic number, which describes how many cuts are needed to disconnect an initially connected graph, falls mostly below $5$e$4$. Due to the clustered values of these four indices, it is not easy to identify the correlation using trend lines, but it can be observed that the highest concentration point appears near the leftest side in IDX~$1,3,4$, and the secondary highest goes right, third right further, and the concentrations continue to decrease as steps for larger values of the indices. If we interpret the distribution using groups, it is easier to find that poorly connected cities have a higher chance of indicating higher concentrations of NO$_2$. 

IDX~$5\text{--}7$ examine the compactness through geodesic distances between any pair of nodes. The higher the system dispersion index (IDX~$7$) or diameter (IDX~$6$), the more dispersed, sprawling, or encroaching the network. The redundancy ratio (IDX~$5$) is inversely proportional to IDX~$7$ for any given $n$; the larger the ratio, the more compact the network. It is found that IDX~$5$ is more sensitive to indicate the NO$_2$ concentrations; when the ratio is greater than 0.05, there is no city with concentrations greater than 20 parts per billion (ppb); when the ratio is greater than 0.09, no city is greater than 10. 

In the last section, we observed three types of geometric patterns that possibly indicate higher concentrations: 1$^{st}$, particles tend to aggregate on walking paths, especially near where there are abrupt shifts from driving paths; 2$^{nd}$, curvy roads; 3$^{rd}$, less regular and more disordered patterns. Therefore, IDX~$8$ is introduced to examine the 1$^{st}$ type of patterns, and IDX~$9$ for the 3$^{rd}$ type. $W_k$ and $D_r$ stand for walking and driving roads, respectively. $D(u,v)$ is the physical distance (not the geodesic distance) between the nodes $u$ and $v$. Most cities have only a maximum 10\% of the total length of paths built for walking, with the IDX~$8$ well below $0.1$. The logarithmic scale plot, as \Cref{fig:fig-emis-p7-p12}\textbf{B}, is used again to fit an extreme city point of Venice in Italy that has almost no driving paths (IDX~$8=15.8$, NO$_2=8.3$~). In addition to that, 10 more global cities have $>50\%$ walking paths, half of which are located in Germany. Cities with NO$_2>20$~ppb have the index $<15.7\%$. 

For the 3$^{rd}$ type, although entropy could be used to indicate the randomness of road orientations \cite{entropy-2019}, the randomness of the network structure is not only about edge orientations. Ricci curvature can reflect whether an edge is within or outside a cluster by comparing the distance between each pair of nodes with the Wasserstein distance $W(m_x,m_y)$. A graph with more edges with positive curvature is more possibly a complete graph, negative for a more dispersed, random, spread-out pattern, and zero-curvature for grid-like patterns\cite{Ni_Lin_Luo_Gao_2019}\cite{Gao_Liu_Liu_Wang_Deng_Zhu_Li_2019}. As in \Cref{fig:fig-emis-p7-p12}\textbf{C}, the distribution indicates that most cities are randomly dispersed structures to which high concentrations of air pollution are more often attached. We did not find a suitable quantity to reflect the 2$^{nd}$ type.

In addition to the observed patterns, two typical networks are also worth looking at from the perspective of air pollution. In the concept of small-world, two physical quantities are measured: characteristic path length and clustering coefficient. The characteristic path length has been examined as the geodesic distance in the IDX~$5\text{--}7$; the clustering coefficient is therefore added as the IDX~$10$ as \Crefrange{eq-10-1}{eq-10-2}. The high clustering coefficient, equal to 1, is found to be correlated with high concentrations as in \Cref{fig:fig-emis-p7-p12}\textbf{D}. 

Following the concept of scale-free models, IDX~$11$ is constructed. Suppose that the degree distribution of the nodes $deg(v)$ for graph $G$ is $<n_0, n_1, \dots, n_{k}>$, where $k$ stands for the degree, $n_k$ for the number of nodes with $k$-degree. If the histogram for$<n_0, n_1, \dots, n_{k}>$ can fit a negative power curve $k$ as \Cref{eq-11-1} (logarithmic \Cref{eq-11-2}), then the network is scale-free. IDX~$11$ is a simplified version of scale-free, which calculates the ratio between the least and most values in $<n_0, n_1, \dots, n_{k}>$. When the two values are close, the ratio approaches $1$, the curve is unlikely to follow the power law, which is not scale-free. As in \Cref{fig:fig-emis-p7-p12}\textbf{E}, most cities could follow a "scale-free" pattern, resulting in higher concentrations. For cities with a ratio approaching $0.7$, the maximum value of this index corresponds to the lowest concentration range. IDX~$12$ as \Cref{eq-12} is used to double-check the connectivity of a graph. Its trend aligns with the trends of IDX~$1\text{--}4$. Most of the cities investigated are confirmed to be weakly connected.

In comparison to geometric patterns discussed in the last section, the 12 indices are mainly used to examine the topological characteristics of transport networks, such as connectivity, clustering, etc. The indices are based on graph theory to quantify the study of topological properties and geometric patterns that we observed on images but were unable to be quantified in the last section. 

\section{Conclusion}
This research aims to understand how transport networks can affect air pollution through varied geometric and topological properties that can be measured with the graph theory tool. The 0.3m images are interpreted with cGAN for pattern recognition, from which the relation between air pollution and the geometric patterns of transport networks is visualised and identified for quantification in combination with topological properties as 12 indices. The sample points of $\sim$ 1700 global cities are plotted to quantify the relationships.

An air-polluted city could suggest some of the following characteristics in its transport network: low connectivity (IDX~$1\text{--}4$, IDX~$12$), low redundancy ratio (IDX~$5$), inadequate paths for walking (IDX~$8$), disordered structure (IDX~$9$), extreme clustering coefficient (IDX~$10$), node-degree potentially following scale-free rule (IDX~$11$). Understandably, we could improve air quality by increasing almost all IDXs except IDX~$10$, where extreme values should be avoided.

If we take a step back from the graph's perspective and return to the transport context, low connectivity, low redundancy ratio, and disordered structure could indicate much longer travel distances than necessary. Inadequate paths for walking and extreme clustering coefficient could resonate with travel behaviour or living community patterns. Node-degree tells the unbalanced degree distribution that could be related to the level of convenience in travel access, leading to continuous busy traffic in some locations while remaining idle in others. Travel distance, behaviour, access, and living community patterns are not new in scientific transport discussion, and many good suggestions have been proposed for environmental concerns. We hope to add one viewpoint using these quantified indices for easier monitoring of the network's impact on air pollution.


%



\ifCLASSOPTIONcompsoc
  \section*{Acknowledgment}
\fi
The author would like to thank Dr.~Haifeng Zhao for technical suggestions on the cGAN model; Prof.~Mark Stevenson and Dr.~Kerry Nice for reviewing the manuscript. The author was funded by the Melbourne Research Scholarship. This research was supported by The University of Melbourne’s Research Computing Services and the Petascale Campus Initiative.
\ifCLASSOPTIONcaptionsoff
  \newpage
\fi



%

\bibliographystyle{IEEEtran} \bibliography{science_template}
%








\end{document}